\documentclass[11pt]{article}

\usepackage{graphicx, amssymb, latexsym, amsfonts, amsmath, lscape, amscd,
amsthm, color, epsfig, mathrsfs, tikz, enumerate}
\usepackage{fullpage}
\usepackage{graphicx}
\usepackage{epstopdf}
\usepackage{float}
\usepackage{caption}
\usepackage{pstricks}
\usepackage{subfig}

\begin{document}

\title{{\bf Hesitancy, Awareness and Vaccination: A Computational Analysis on Complex Networks }}
\author{
{\sc Dibyajyoti Mallick*}, {\sc Aniruddha Ray*}, {\sc Ankita Das*}, {\sc Sayantari Ghosh}\\
\mbox{}\\
{\small Department of Physics, National Institute of Technology, Durgapur}\\
{\small* These authors share equal credit in this work.}}

\date{}

\maketitle

\begin{abstract}

Considering the global pandemic of coronavirus disease 2019 (COVID-19), around the world several vaccines are being developed. Till now, these vaccines are the most effective way to reduce the high burden on the global health infrastructure.  However, the public acceptance towards vaccination is a crucial and pressing problem for health authorities. This study has been designed to determine the parameters affecting the decisions of common individuals towards COVID-19 vaccine.  In our study, using the platforms of compartmental model and network simulation, we categorize people and observe their motivation towards vaccination in a mathematical social contagion process.  In our model, we consider peer influence as an important factor in this dynamics, and study how individuals are influencing each other for vaccination. The efficiency of the vaccination process is estimated by the period of time required to vaccinate a substantial fraction of total population. We discovered the major barriers and drivers of this dynamics, and concluded that it is required to formulate specific strategies by the healthcare workers which could be more effective for the undecided and vaccine hesitant group of people.
\end{abstract}

\section{Introduction}
The novel coronavirus disease 2019(COVID-19), caused by severe acute respiratory syndrome coronavirus 2 (SARS-CoV-2), was first reported in Wuhan, Hubei Province of China. This disease is now a global pandemic which is spreading rapidly from person to person causing major public health concerns and economic crisis \textcolor{blue}{\cite{nishi2020network}} \textcolor{blue}{\cite{world2020economy}}. A variety of active intervention policies have been introduced to suppress the spreading of this disease, such as hand sanitizing, social distancing, travel restrictions, partial or complete lockdown, wearing mask, quarantining, etc. After the declaration of the pandemic by WHO (World Health Organization) in March 2020, pharmaceutical companies and scientists have encountered a race against time to develop vaccines \textcolor{blue}{\cite{habersaat2020understanding}}. \\
The recent availability of multiple vaccines against coronavirus has brought hope for prevention of the spreading and a rapid recovery of our badly affected economy, with a promise of sooner resumption of normal life. However, the widespread hesitancy about vaccines are becoming a major obstacle for global health. Many people have strong hesitation towards vaccination which is defined as the confusion about safety and effectiveness of the vaccine. The origin of this usually lies in some rumors regarding vaccine, concerning about side effects. This is has become a huge challenge for governments and public health authorities for reaching the expected and required vaccination coverage \textcolor{blue}{\cite{dror2020vaccine}} \textcolor{blue}{\cite{kumar2016vaccine},\cite{troiano2021vaccine}}. The key priority now is ensuring the vaccine acceptance because the lag in vaccination may provide a window for spread the new variants and can also be a major obstacle in developing society-wide herd immunity.\\
Till date many researchers, scientists have tried to conduct different surveys to understand the behaviour towards vaccine acceptance and hesitancy \textcolor{blue}{\cite{kalimeri2019human,larson2015measuring, bhattacharyya2019impact}}.  Some studies \textcolor{blue}{\cite{jarrett2015strategies},} have applied different theoretical models to explain the vaccine acceptance, hesitancy, willingness of individuals towards vaccine as well as refusal to vaccinate which may vary depending upon personal decisions and epidemiological conditions \textcolor{blue}{\cite{siddiqui2013epidemiology}}.  Many advance countries are trying to conduct country specific surveys \textcolor{blue}{\cite{lazarus2021global}}, while  several reports are being prepared by different countries such as United Kingdom(UK), United States of America(USA), China, India, Saudi Arab to understand this vaccine acceptance behaviour. \textcolor{blue}{\cite{eskola2015deal} \cite{pronyk2019vaccine} \cite{luyten2019assessing}}. \\
However, most of this studies are based on heuristic arguments rather than mathematical analysis. A computational framework having mathematical foundation helps to draw quantitative conclusions, and much more effective in predictive modeling purposes. This paper represents a new mathematical model regarding vaccination process which incorporates behavioural changes of every individuals in a society, driven by the global and local factors. The main assumption is here that the vaccination dynamics can be considered as social contagion process. This study aimed to identify the acceptability of covid-19 vaccine, information in support or opposing vaccination are flowing in a society like a viral infection. We take into account of refusal and hesitancy factors as well as the positive attitude of people towards vaccination \textcolor{blue}{\cite{sallam2021covid}}. In our study we focused on to identify the effects of the factors that could increase the vaccination coverage through numerical simulations on an artificial society. Moreover, we could justify several results found by survey based studies where this kind of questions have already been explored using survey results \textcolor{blue}{\cite{dube2015vaccine, peretti2015vaccine}}. Our results address the vaccination acceptance problem using a comprehensive mathematical and computational framework, that would help us figuring out correct strategies, to encourage community for vaccine uptake and to stop further spreading of this pandemic.

\section{Model Formulation}\label{modelform}

To understand this kind of problem, we have considered a set of differential equations to depict the possible transitions. To describe the vaccine dynamics we have our compartmental model as shown in figure 1 , at any time $t$, the total population $N(t)$ is subdivided into four states: Ignorant ($I$), Hesitant ($H$), Unwilling($U$) , Vaccinated($V$).

\begin{itemize}
    \item Ignorant($I$) $-$  As the name suggests, in this group, people are ignorant about vaccine. They have no idea about vaccination, and have no clear opinion. In general, these people will not involve themselves in vaccination process. But they may be influenced by hesitant and vaccinated groups as well. Each people in this group are represented by $I(t)$.
    \item Hesitant($H$) $-$ In this group people know about vaccine nevertheless, they feel hesitation while taking decision regarding vaccination, being affected by the rumors. This group may be influenced by the global advertisement campaigns in favour of vaccination, and take vaccine eventually. On the other hand, they could also be influenced further by rumors and transit to unwilling state. Each people in this group are represented by $H(t)$. 
    \item Unwilling($U$) $-$ In this group people believe on negative rumors regarding vaccine and they push themselves for vaccination. Again some Unwilling people may return to Hesitant class as far they can not decided whether to get vaccine or not, by getting influenced through vaccinated subpopulation.  Each people in this group are represented by $U(t)$.
    \item Vaccinated($V$) $-$ In this group people are completely vaccinated and they are trying to influence other groups for joining them. Each people in this group are represented by $V(t)$.
    \end{itemize}
In this model the key variable we have is the vaccinated population, who are influencing other populations to get vaccinated. This is also the target population which have to maximised over a certain period of time. Here, this group of people is playing an effective role to control the epidemic, and spread the infodemic.\\
 Considering the total population = 1 (normalized form),  
 \begin{equation*}
     I+H+U+V = 1
 \end{equation*}
Now, we will discuss the following possibilities of transition of people from one subpopulation to another over a given time period `t'. Let us discuss all the possible transitions considered in model, categorically.
\begin{figure}
    \centering
    \includegraphics[width=0.8\linewidth]{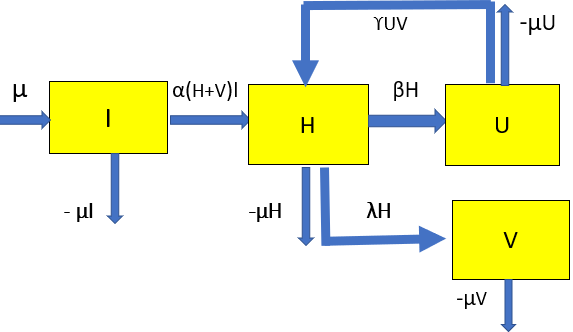}
    \caption{{Compartmental diagram of the proposed model ($I$-$H$-$U$-$V$). All the transitions, considered in the proposed model are shown through blue arrows. Subpopulations are denoted by yellow boxes.}}
    \label{fig:1}
\end{figure}
\begin{itemize}
    \item {Possible transitions related to $I$ subpopulation:}\\In this population $\mu$ is the rate at which people are entering into the ignorant population. This is added to consider the demographic variations. Now the members of $I$ may be influenced by $H$ and $V$ population. $H$ and $V$ people might spread words on their inclination and decision, and encourage others for vaccination. Let us consider $\alpha$ be the effective contact rate. If each member of $I$ group is going to be influenced by hesitant and vaccinated group with that rate, then $\alpha (H+V) I$ amount will be subtracted from $I$ group and must be added to hesitant group. In the normal course, there must be some people who can die from that population, hence the term $\mu$ should also be deducted from group $I$ in the rate equation.
  
\item {Possible transitions related to $H$ subpopulation:}\\
As mentioned in the previous case, due to the effect of influence from group $H$ and $V$, the term $\alpha (H+V)$ should be added in this rate equation. As usual people from this group can also die hence the term $\mu H$ also be deducted from $H$ group in its rate equation. Due to some reasons (influencing from other groups or due to development of self awareness among people), they are going for vaccination at a rate $\lambda$ and some are entering into Unwilling group at a rate $\beta$. Hence the term $\beta H$ and $\lambda H$ should be deducted from the rate equation. Again from Unwilling group some people are coming back into Hesitant group by believing on some rumors, hence the term $\gamma U V$ should be added to the rate equation of $H$.
\item {Possible transitions related to $U$ subpopulation:}\\
From the previous case discussed above, it should be noted that the term $\gamma U V$ and $\mu U$ should be deducted from its rate equation, due to their hesitant behaviour and natural death from this group respectively. 

\item {Possible transitions related to $V$ subpopulation:}\\From the discussion of case 2, it is evident that the term $\lambda H$ must be added to its rate equation due to direct wishing for vaccination from group $H$ and $\mu V$ must be deducted due to natural death of people from this group.
\end{itemize}
Hence all the rate equations of this model are compiled as follows:
\begin{align}
\frac{dI}{dt}=\mu-\alpha(H+V)I-\mu I
\end{align}
\begin{align}
\frac{dH}{dt}=\alpha(H+V)I-\beta H-\lambda H+{\gamma}\; UV-\mu H
\end{align}
\begin{align}
\frac{dU}{dt}=\beta H-\gamma \; UV-\mu U
\end{align}
\begin{align}
\frac{dV}{dt}=\lambda H-\mu V
\end{align}
\section{Model Analysis on Complex Network}\label{results}
In ODE-based models one of the major issues are homogeneous mixing, which indicates every individuals in a population have same probability of having contact with each other \textcolor{blue}{\cite{wang2020evaluation}}. Our society in highly heterogeneous, and to accommodate that fact into our findings, we study the model on the heterogeneous setting of a complex network. The simulations are performed on a random network having 10000 nodes with an average degree 5. Here we choose EoN module in Networkx from python \textcolor{blue}{\cite{miller2020eon}} and run these following simulations in Google Colaboratory. Varying the transition rates of each parameter we see the effects of that particular parameter on the time evolution of the dynamics through our model.\\
The general dynamics has been depicted in Figure 2. It shows that eventually most of the people in population get vaccinated, however the time of coverage might be different, depending on the parameter values. Thus, to quantify this growth curve, we define, vaccination coverage time, $\tau_V$:
\begin{equation*}
    \tau_V=\{t|V(t)=0.9\;V_{max}\}
\end{equation*}
This means that we consider a timescale, $\tau_V$ within which 90\% of maximum vaccination, $V_{max}$ has been achieved. This will give us an estimate of the vaccination coverage speed.
\begin{figure}
    \centering
    \includegraphics[width=0.6\linewidth]{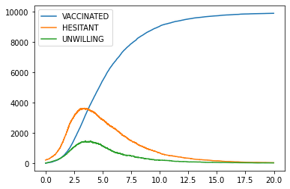}
    \caption {Population of different classes vs. time. Eventually most of the population gets vaccinated, but time of coverage depends on parameters.}
    \label{fig:2}  
\end{figure}

\subsection{Effect of Vaccination Rate}
\begin{figure}
    \centering
    \includegraphics[width=\linewidth]{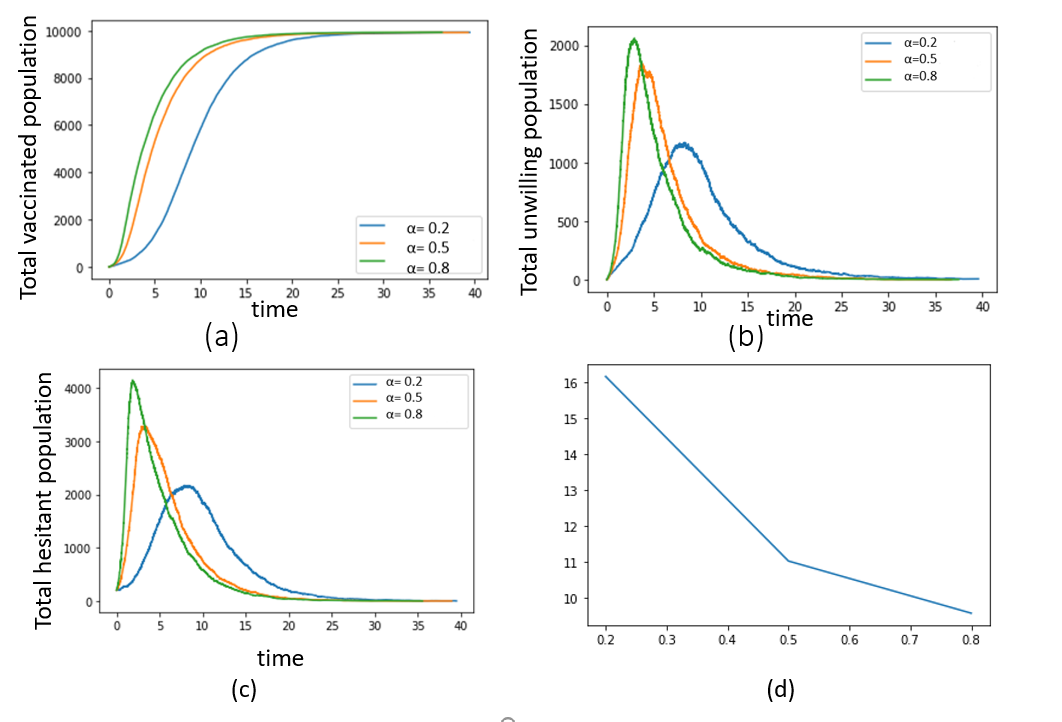}
    \caption{{(a) Vaccination coverage with time for different values of vaccination rate $\lambda$}. (b) Time evolution of Unwilling people in the population with different values of vaccination rate parameter. (c) Time evolution of Hesitant people in the population with different values of vaccination rate parameter. (d) Vaccination coverage time($\tau_V$) for different values of vaccination rate.}
    \label{fig:3}
\end{figure}

Vaccination rate is a major parameter of this dynamics. If you carefully observe different sub-populations, the effect could be prominently observed, as reported in Figure \ref{fig:3}.

\begin{itemize}
    \item Effect on vaccinated people: As we can see from the results in Figure 3(a), that increasing the rate does not change the maximum number of vaccinated people but it shifts the saturation point leftwards, which means the system is reaching the saturation point faster.
    \item Effect on hesitant people: We can see from the plots shown in Figure 3(b) that increasing the rate shifts the peak point slightly leftwards, which means the system is reaching the peak point a bit faster and we also can see that the maximum number of people reaching in a hesitant class decreases a little bit.
    \item Effect on unwilling people: As we can see from the plots shown in Figure 3(c), increasing the parameter flattens the curve which means that the maximum number of people going to the unwilling class decreases that can be explained by the success rate of vaccination process.
    \end{itemize}
  As in figure 3(d), we can see that the saturation time decreases with increasing vaccination rate that is, $\lambda$ which states that this particular parameter will make the system to reach the saturation point earlier.
\subsection{Effect of Negative Rumors}  
\begin{figure}
    \centering
    \includegraphics[width=1.0\linewidth]{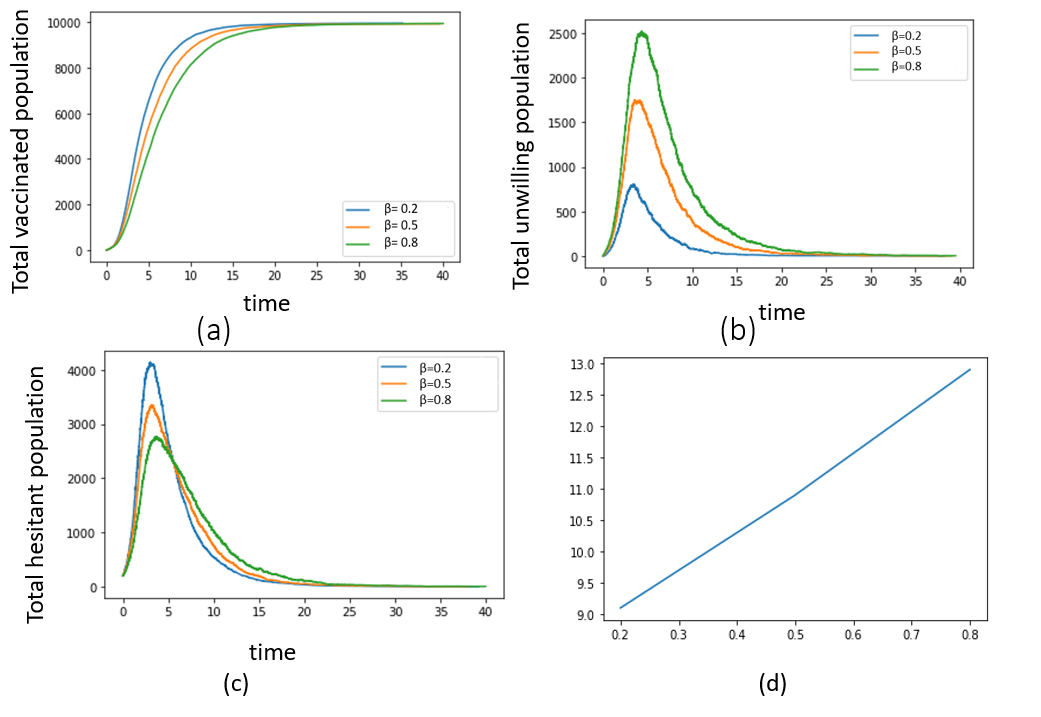}
    \caption{{(a) Vaccination coverage with time for different values of negative rumor influence rate $\beta$}. (b) Time evolution of Unwilling people in the population with different values of negative rumor influence rate parameter. (c) Time evolution of Hesitant people in the population with different values of negative rumor influence rate parameter. (d) Vaccination coverage time($\tau_V$) for different values of negative rumor influence rate.}
    \label{fig:4}
\end{figure}
Negative rumors often sway people towards some risky and dangerous activity. In this vaccination process some negative rumors about vaccine exist that concerns with matters like, several side effects, cost effective, fear, misinformation regarding vaccine. Hesitant people often are getting influenced by them. The parameter that takes into account of this phenomena is $\beta$. The results related to this are shown in Figure 4. 
\begin{itemize}
     \item Effect on vaccinated class:  As we can see in Figure 4(a), increasing the parameter, $\beta$, shifts the saturation point of vaccinated people rightwards. This means as more people feel negative about vaccination, more time it will take for the system to reach the saturation point. 
     \item Effect on unwilling class: In Figure 4(b) we can see that many people are motivated from their neighbors, friends, relatives to believe on some negative rumors about vaccination and denied to get vaccine and along the way the maximum number of unwilling people increases.
     \end{itemize}
As in figure 4(d) we can see that the saturation time increases with increasing the negative peer influence rate that is, $\beta$. This means the system will take more time to reach the saturation point.

\subsection{Effect of Positive Peer Influence}
\begin{figure}
    \centering
    \includegraphics[width=1.0\linewidth]{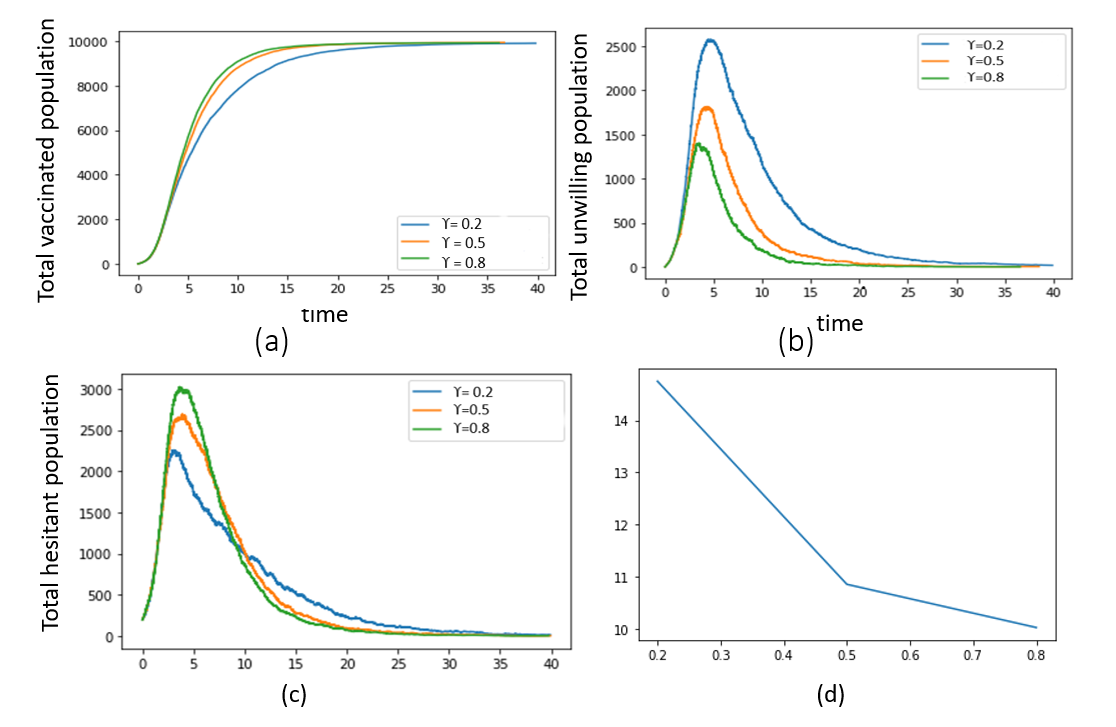}
    \caption{{(a) Vaccination coverage with time for different values of positive peer influence rate $\gamma$}. (b) Time evolution of Unwilling people in the population with different values of positive peer influence rate parameter. (c) Time evolution of Hesitant people in the population with different values of positive peer influence rate parameter. (d) Vaccination coverage time($\tau_V$) for different values of positive peer influence rate.}
    \label{fig:5}
\end{figure}
\begin{figure}
    \centering
    \includegraphics[width=1.0\linewidth]{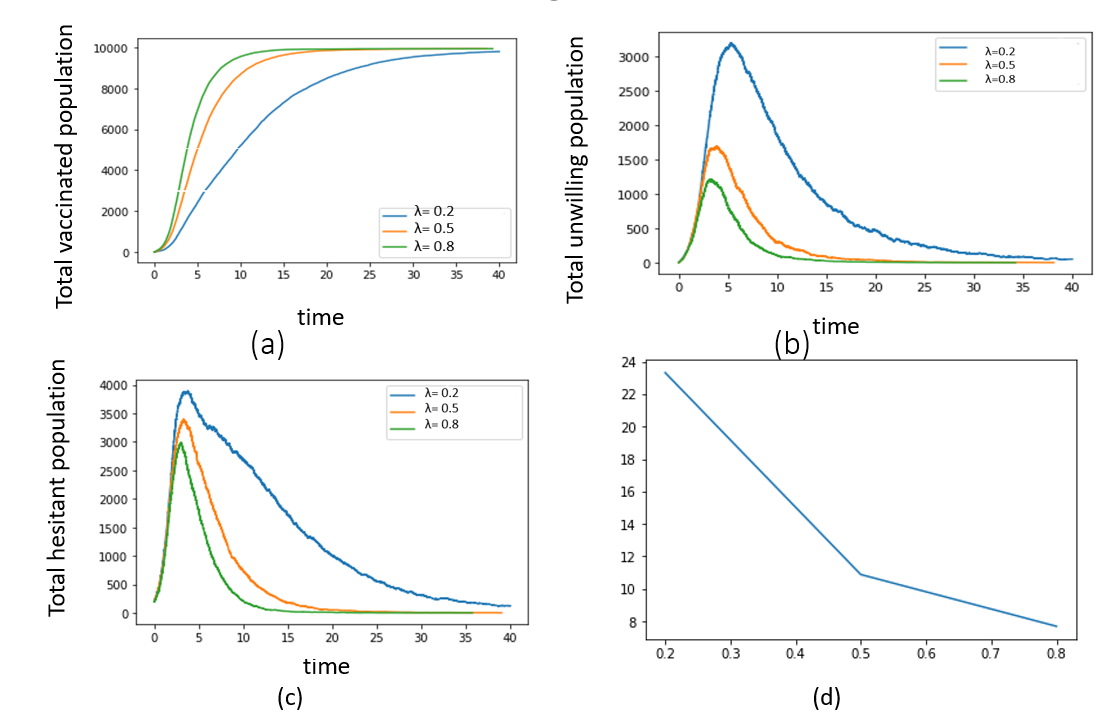}
    \caption{{(a) Vaccination coverage with time for different values of positive peer influence rate $\alpha$}. (b) Time evolution of Unwilling people in the population with different values of positive peer influence rate $\alpha$. (c) Time evolution of Hesitant people in the population with different values of positive peer influence rate $\alpha$. (d) Vaccination coverage time($\tau_V$) for different values of positive peer influence rate $\alpha$.}
    \label{fig:6}
\end{figure}
Peer influence occurs when people are motivated or influenced towards something by seeing their neighbour or friends. When someone's peers influence them to do something positive is considered as positive peer influence. In the vaccination context, positive peer influence occurs when someone who is vaccinated and he/she is motivating others who are not sure about vaccination. Here the vaccinated people influences the hesitant people to get vaccinated. Here we are considering two parameters $\alpha$ and $\gamma$ that takes into account this phenomena. The results related to this are shown in Figure 5 and Figure 6.

 \begin{itemize}
     \item Effect on Vaccinated people: As we can see in Figure 5(a) and 6(a), that increasing the parameter it shifts the saturation point of vaccinated people leftwards that means the more people feel positive about vaccination  then sooner it will take the system to reach the saturation point. 
     \item Effect on Unwilling people: We can also see from Figure 5(b) and 6(b) that along the way the maximum number of unwilling people decreases as they are getting motivated by some positive rumors towards vaccination.
     \end{itemize}
So, we can draw this conclusion that if more people are well informed about the success rate of vaccination then that would fasten up the vaccination process and also decrease the amount of unwilling people. \\
As shown in figure 5(d) and 6(d), for both parameters of positive peer influence that is, $\alpha$ and $\gamma$, the saturation time decreases with increasing parameters which means positive peer influence will make the system reach the saturation point earlier.

\section{Conclusions and perspectives} \label{section:conclusion}
Along with the infectious spread of SARS-COV2, the information and misinformation regarding available vaccines are also spreading person-to-person causing a large-scale effect on the vaccination coverage. Through computational analysis and network simulations, we have explored this contagion dynamics, and proposed a framework to analyze this decision process. From the model it has been observed that how the presence of the efficient vaccination system can run the entire dynamics with positive feedback. This is a significant result as we can see a lot of people not only in India even around the world are hesitant about getting vaccinated.\\
In our model, we have shown that if the peer influence is overall positive then the saturation point shifts leftwards, so we can conclude that the  spreading of awareness about the positive effects of vaccination should be carried out on the ground level. We have also shown in our model that if vaccination rate increases then the saturation point of $V$ class shift leftwards, in real life we can explain it like this$-$ if the vaccination rate in a particular area can be increased by awareness and smooth execution of the vaccination process by competent authority then the system would reach its saturation point sooner. A strong peer influence also might signify a strongly connected society. The shift the saturation point leftwards for high positive peer influence means that physically and virtually everyone is well connected. This may also indicate high vaccination rates in urban areas. \\
In future scenario, we will focus on implementing further realistic terms in our model. For example, if someone from hesitant class goes to unwilling class it can be considered that the person was under the influence of someone who has been vaccinated, thus we can bring the non-linearity. Mathematical explorations to find out The fixed points and bifurcations if there are any. Considering coupled dynamics, or delayed dynamics of disease along with the infodemic might be another interesting future study.

\bibliographystyle{abbrv}
\bibliography{bibliography.bib}

\begin{thebibliography}{10}

\bibitem{world2020economy}
W.~Bank.
\newblock The economy in the time of covid-19, 2020.

\bibitem{bhattacharyya2019impact}
S.~Bhattacharyya, A.~Vutha, and C.~T. Bauch.
\newblock The impact of rare but severe vaccine adverse events on
  behaviour-disease dynamics: a network model.
\newblock {\em Scientific reports}, 9(1):1--13, 2019.

\bibitem{dror2020vaccine}
A.~A. Dror, N.~Eisenbach, S.~Taiber, N.~G. Morozov, M.~Mizrachi, A.~Zigron,
  S.~Srouji, and E.~Sela.
\newblock Vaccine hesitancy: the next challenge in the fight against covid-19.
\newblock {\em European journal of epidemiology}, 35(8):775--779, 2020.

\bibitem{dube2015vaccine}
E.~Dub{\'e}, M.~Vivion, and N.~E. MacDonald.
\newblock Vaccine hesitancy, vaccine refusal and the anti-vaccine movement:
  influence, impact and implications.
\newblock {\em Expert review of vaccines}, 14(1):99--117, 2015.

\bibitem{eskola2015deal}
J.~Eskola, P.~Duclos, M.~Schuster, N.~E. MacDonald, et~al.
\newblock How to deal with vaccine hesitancy?
\newblock {\em Vaccine}, 33(34):4215--4217, 2015.

\bibitem{habersaat2020understanding}
K.~B. Habersaat and C.~Jackson.
\newblock Understanding vaccine acceptance and demand—and ways to increase
  them.
\newblock {\em Bundesgesundheitsblatt-Gesundheitsforschung-Gesundheitsschutz},
  63(1):32--39, 2020.

\bibitem{jarrett2015strategies}
C.~Jarrett, R.~Wilson, M.~O’Leary, E.~Eckersberger, H.~J. Larson, et~al.
\newblock Strategies for addressing vaccine hesitancy--a systematic review.
\newblock {\em Vaccine}, 33(34):4180--4190, 2015.

\bibitem{kalimeri2019human}
K.~Kalimeri, M.~G.~Beir{\'o}, A.~Urbinati, A.~Bonanomi, A.~Rosina, and
  C.~Cattuto.
\newblock Human values and attitudes towards vaccination in social media.
\newblock In {\em Companion Proceedings of The 2019 World Wide Web Conference},
  pages 248--254, 2019.

\bibitem{kumar2016vaccine}
D.~Kumar, R.~Chandra, M.~Mathur, S.~Samdariya, and N.~Kapoor.
\newblock Vaccine hesitancy: understanding better to address better.
\newblock {\em Israel journal of health policy research}, 5(1):1--8, 2016.

\bibitem{larson2015measuring}
H.~J. Larson, C.~Jarrett, W.~S. Schulz, M.~Chaudhuri, Y.~Zhou, E.~Dube,
  M.~Schuster, N.~E. MacDonald, R.~Wilson, et~al.
\newblock Measuring vaccine hesitancy: the development of a survey tool.
\newblock {\em Vaccine}, 33(34):4165--4175, 2015.

\bibitem{lazarus2021global}
J.~V. Lazarus, S.~C. Ratzan, A.~Palayew, L.~O. Gostin, H.~J. Larson, K.~Rabin,
  S.~Kimball, and A.~El-Mohandes.
\newblock A global survey of potential acceptance of a covid-19 vaccine.
\newblock {\em Nature medicine}, 27(2):225--228, 2021.

\bibitem{luyten2019assessing}
J.~Luyten, L.~Bruyneel, and A.~J. van Hoek.
\newblock Assessing vaccine hesitancy in the uk population using a generalized
  vaccine hesitancy survey instrument.
\newblock {\em Vaccine}, 37(18):2494--2501, 2019.

\bibitem{miller2020eon}
J.~C. Miller and T.~Ting.
\newblock Eon (epidemics on networks): a fast, flexible python package for
  simulation, analytic approximation, and analysis of epidemics on networks.
\newblock {\em arXiv preprint arXiv:2001.02436}, 2020.

\bibitem{nishi2020network}
A.~Nishi, G.~Dewey, A.~Endo, S.~Neman, S.~K. Iwamoto, M.~Y. Ni, Y.~Tsugawa,
  G.~Iosifidis, J.~D. Smith, and S.~D. Young.
\newblock Network interventions for managing the covid-19 pandemic and
  sustaining economy.
\newblock {\em Proceedings of the National Academy of Sciences},
  117(48):30285--30294, 2020.

\bibitem{peretti2015vaccine}
P.~Peretti-Watel, H.~J. Larson, J.~K. Ward, W.~S. Schulz, and P.~Verger.
\newblock Vaccine hesitancy: clarifying a theoretical framework for an
  ambiguous notion.
\newblock {\em PLoS currents}, 7, 2015.

\bibitem{pronyk2019vaccine}
P.~Pronyk, A.~Sugihantono, V.~Sitohang, T.~Moran, S.~Kadandale, S.~Muller,
  C.~Whetham, and R.~Kezaala.
\newblock Vaccine hesitancy in indonesia.
\newblock {\em The Lancet Planetary Health}, 3(3):e114--e115, 2019.

\bibitem{sallam2021covid}
M.~Sallam.
\newblock Covid-19 vaccine hesitancy worldwide: a concise systematic review of
  vaccine acceptance rates.
\newblock {\em Vaccines}, 9(2):160, 2021.

\bibitem{siddiqui2013epidemiology}
M.~Siddiqui, D.~A. Salmon, and S.~B. Omer.
\newblock Epidemiology of vaccine hesitancy in the united states.
\newblock {\em Human vaccines \& immunotherapeutics}, 9(12):2643--2648, 2013.

\bibitem{troiano2021vaccine}
G.~Troiano and A.~Nardi.
\newblock Vaccine hesitancy in the era of covid-19.
\newblock {\em Public Health}, 2021.

\bibitem{wang2020evaluation}
N.~Wang, Y.~Fu, H.~Zhang, and H.~Shi.
\newblock An evaluation of mathematical models for the outbreak of covid-19.
\newblock {\em Precision Clinical Medicine}, 3(2):85--93, 2020.

\end{thebibliography}

\end{document}